\documentclass[conference]{IEEEtran}
\IEEEoverridecommandlockouts
\usepackage{cite}
\usepackage{amsmath,amssymb,amsfonts}
\usepackage{algorithmic}
\usepackage{graphicx}
\usepackage{textcomp}
\usepackage{xcolor}
\usepackage{hyperref}

\usepackage{multicol}
\begin{document}

\title{Assessing Risks and Modeling Threats in the Internet of Things}

\author{\IEEEauthorblockN{Paul Griffioen}
\IEEEauthorblockA{Department of Electrical and Computer Engineering \\
Carnegie Mellon University \\
Pittsburgh, PA, USA \\
pgriffi1@andrew.cmu.edu}
\and
\IEEEauthorblockN{Bruno Sinopoli}
\IEEEauthorblockA{Department of Electrical and Systems Engineering \\
Washington University in St. Louis \\
St. Louis, MO, USA \\
bsinopoli@wustl.edu}
}

\maketitle

\begin{abstract}
Threat modeling and risk assessments are common ways to identify, estimate, and prioritize risk to national, organizational, and individual operations and assets. Several threat modeling and risk assessment approaches have been proposed prior to the advent of the Internet of Things (IoT) that focus on threats and risks in information technology (IT). Due to shortcomings in these approaches and the fact that there are significant differences between the IoT and IT, we synthesize and adapt these approaches to provide a threat modeling framework that focuses on threats and risks in the IoT. In doing so, we develop an IoT attack taxonomy that describes the adversarial assets, adversarial actions, exploitable vulnerabilities, and compromised properties that are components of any IoT attack. We use this IoT attack taxonomy as the foundation for designing a joint risk assessment and maturity assessment framework that is implemented as an interactive online tool. The assessment framework this tool encodes provides organizations with specific recommendations about where resources should be devoted to mitigate risk. The usefulness of this IoT framework is highlighted by case study implementations in the context of multiple industrial manufacturing companies, and the interactive implementation of this framework is available at \url{http://iotrisk.andrew.cmu.edu}.
\end{abstract}

\section{Introduction}
An increasing number of everyday devices are being connected to the Internet, creating a network of physical devices, home appliances, and other items embedded with unique sets of electronics, software, sensors, and actuators. Forecasts predict that by 2025, about 73 billion Internet of Things (IoT) devices will be employed throughout the world \cite{IHS}. This rapid growth in the IoT creates opportunities for more direct integration of the physical world into computer-based systems but also introduces a plethora of risks. IoT devices, whether they be part of smart transportation systems, thermostats that adapt to daily lifestyles, or medical devices that can monitor a patient in real time, are not always built with security and privacy in mind, leaving them extremely vulnerable to attacks.

An example of a pervasive IoT attack was the Mirai botnet, where an attacker gained unauthorized access to numerous IoT devices including IP cameras and older routers. These infected devices were used to make much of the Internet unavailable by overwhelming Dyn, a domain name system (DNS) provider. The malicious code took advantage of the fact that most users do not change the default usernames and passwords on their devices \cite{Mirai}. In addition to this example, there have been other instances of IoT botnets \cite{IoTBotnet} including Persirai \cite{Persirai}, Hajime \cite{Hajime}, BrickerBot \cite{BrickerBot}, and one targeting a university \cite{Verizon}.

Other IoT devices are also susceptible to attacks, as researchers demonstrated that St. Jude Medical pacemakers can easily be compromised from up to 50 feet away, allowing attackers to administer inappropriate pacing or shocks or causing rapid battery depletion \cite{MuddyWaters}. Other medical IoT devices can also be easily compromised including insulin pumps \cite{InsulinBlackHat} and medical infusion pumps \cite{InfusionFDA}. In addition, smart locks \cite{DEFCON}, smart lights \cite{PhilipsHue}, security cameras \cite{DevilsIvy}, smart TVs \cite{SmartTV}, and smart toys \cite{Cloudpets} have been shown to be susceptible to attacks.

The susceptibility of IoT devices to these types of attacks is largely due to a few factors that distinguish the IoT from other areas such as information technology (IT) \cite{NISTNoT,NISTIoTTrust,NISTManageIoT}. One factor is the presence of unfixable flaws. Since the lifetime of many IoT devices is quite large, devices containing vulnerabilities or flaws will continue to be deployed and used long after vendors cease to produce or support them. Secondly, the diversity found across IoT devices and the settings in which they are used means that traditional approaches of discovering attack signatures is insufficient and unscalable. Lastly, the implicit and explicit complex interactions between IoT devices allow attackers to leverage cross-device dependencies, couplings, and dynamic environments to increase the attack space \cite{IoTFlaws}.

Because these attacks can compromise large numbers of IoT devices with ease and because many factors that contribute to IoT device susceptibility are unique to the IoT, organizations must assess threats and risks in IoT environments with IoT-specific frameworks and tools. Organizations may play the roles of IoT consumers, IoT producers, and/or platform operators that provide software to which customers connect their devices. Regardless of the role, each organization must assess and respond to its own IoT security and privacy risks, deciding which actions or controls to take to mitigate those risks. Consequently, this paper firstly synthesizes and adapts previous threat models to design an IoT-specific threat modeling framework. Secondly, the IoT-specific threat modeling framework is leveraged to develop an online interactive tool (\url{http://iotrisk.andrew.cmu.edu}) which implements a joint risk and maturity assessment framework that provides organizations with automated, quantitative analysis and recommendations about how to mitigate risk in their IoT ecosystems. Lastly, the usefulness of this interactive tool and IoT framework is demonstrated by analyzing multiple industrial manufacturing companies.

The remainder of the paper is organized as follows. Section II introduces previous threat models, their shortcomings, and how they can be synthesized and supplemented with an IoT attack taxonomy to effectively model threats in the IoT. Section III presents an overview of the entire IoT framework used to analyze organizations' IoT-specific risk and demonstrates where the interactive tool fits into that framework. Section IV describes the implementation of the interactive tool in leveraging the threat modeling framework to conduct risk and maturity assessments. Section V applies this framework and tool to multiple industrial manufacturing companies, and Section VI concludes the paper.

\section{Synthesizing Prior Threat Models with the IoT}

\subsection{Previous Threat, Risk, and Maturity Models}
Currently, a variety of threat modeling, risk assessment, and maturity assessment frameworks are used by organizations to describe potential threats and indicate areas of improvement, each of which is centered around the measure of risk.\vspace{0.1cm}

\noindent\textit{1) Current Approaches}\\\indent
Risk is a measure of the extent to which an entity is threatened by a circumstance or event, and it is a function of the likelihood of the event and the adverse impact caused by the event. Risk assessment is commonly seen as a means to identify, estimate, and prioritize risk to national, organizational, and individual operations and assets. The traditional approach to risk assessment has been to identify relevant threats to organizations, internal and external vulnerabilities to organizations, the impact or harm to organizations that may occur if vulnerabilities are exploited, and the likelihood that harm will occur due to a successful attack \cite{NISTRiskAssessment}. After the identified risks have been prioritized, appropriate and effective actions and controls are chosen that mitigate those risks.
Several popular and well-regarded approaches to threat modeling and risk assessment include STRIDE \cite{STRIDE}, PASTA \cite{PASTA}, Trike \cite{Trike}, NIST SP-800-30 \cite{NISTRiskAssessment}, ISO/IEC 27005 \cite{ISO27005}, ISO/IEC 31010 \cite{IEC31010}, CRAMM \cite{CRAMM}, FRAP \cite{FRAP}, COBRA \cite{COBRA}, CORAS \cite{CORAS}, and OCTAVE \cite{OCTAVE}.

Throughout the threat modeling and risk assessment process that each of these approaches take, a variety of risk factors and the relationships between each of the factors are assessed to determine levels of risk. Typical risk factors include threats, vulnerabilities, likelihood, impact, and predisposing conditions. The threat modeling process characterizes the various adversarial actions or threats that can adversely impact organizational operations and assets. Identifying vulnerabilities includes finding weaknesses in systems, security procedures, internal controls, or implementation that can be exploited by a threat source. By taking into account adversary intent, capability, and targeting, likelihood denotes the probability that a given threat is capable of exploiting a specific vulnerability. Impact, which is partly determined by identifying high-value assets to the organization, represents the magnitude of harm that would result from a specific threat being carried out. Predisposing conditions are conditions that exist within an organization that may increase the likelihood of attacks causing adverse impacts to organizational operations and assets \cite{NISTRiskAssessment,ISO27005,PwCRisk}. Each of these factors are used in threat modeling and risk assessments in a variety of ways, including quantitatively, qualitatively, or semi-qualitatively \cite{NISTRiskAssessment,ISO27005}. Regardless of which approach is utilized, risk frameworks usually analyze risk by centering and developing around a particular starting point, including threat-oriented, impact-oriented, or vulnerability-oriented approaches \cite{NISTRiskAssessment,PwCRisk}.

To mitigate the risks prevalent in a particular organization, security and privacy controls are implemented as safeguards or countermeasures to prevent and protect against threats. Maturity assessments are used to compile and evaluate the information needed by organizational officials in determining how effective security and privacy controls are in mitigating risks to organizational operations and assets. Maturity assessments are typically comprised of a set of assessment objectives, methods, and objects, and they provide the extent to which controls are implemented correctly, operating as intended, and meeting the security and privacy requirements of the organization \cite{NISTMaturityAssessment}.\vspace{0.1cm}

\noindent\textit{2) Shortcomings of Current Approaches}\\\indent
Since existing threat modeling and risk assessment methodologies were established prior to the development of the IoT, they are ill-equipped to effectively address IoT-specific threats and vulnerabilities. Existing frameworks only focus on individual assets, devices, and communication platforms, but in the IoT, it is necessary for frameworks to thoroughly consider the relationships, processes, and couplings between IoT devices that arise from the complex and pervasive nature of IoT ecosystems. For example, an attacker could compromise a smart plug to turn off the air conditioning, triggering a temperature increase. If there is a connected service that opens the windows when the temperature rises above a certain level, then this temperature increase may cause the windows to open, resulting in a physical security breach. However, existing frameworks focus on individual devices and do not consider these interactions where the response of one IoT device serves as the input to other IoT actors \cite{IoTRiskAssessment}.

In addition, existing frameworks implement periodic assessments which make the faulty assumption that systems do not change significantly in short periods of time \cite{TaubenbergerIT}. This assumption does not hold well in the IoT where there is much variability in system scale, dynamics, and coupling, implying that there is a high probability of a new system emerging between periodic assessments. Current frameworks also have a simplistic view of organizational assets, only seeing them as things of value. In the IoT, however, organizational assets are more than just things of value and can be platforms from which attacks are launched as in the case of the Mirai botnet \cite{IoTRiskAssessment}.

\begin{figure}[h]
\centering
\includegraphics[width=\linewidth]{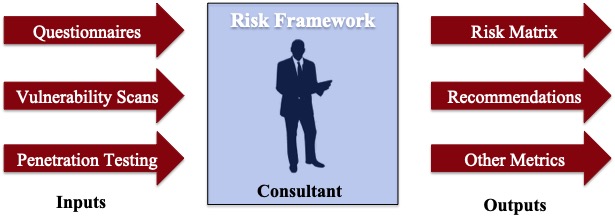}
\caption{Current Threat Modeling and Risk Assessment Methodologies}
\label{fig:Consultant1}
\end{figure}

Another shortcoming of current threat modeling and risk assessment methodologies is not taking into account the distinctive features that render IoT devices susceptible to attacks, including unfixable flaws, a wide range of diversity, complex interactions between devices, and a direct impact on physical processes. Existing frameworks, including the most popular approaches such as \cite{NISTRiskAssessment,ISO27001,OCTAVE}, are typically qualitative rather than quantitative and depend on subjective input from consultants. As depicted in Figure \ref{fig:Consultant1}, the consultant is usually heavily involved in the computation and analysis of the threats and risks as opposed to simply gathering relevant information to input to an automated framework. This results in a lack of precision and much variability between recommendations provided by different consultants. For example, one person's view of a threat being low (as opposed to medium or high) might not conform to another person's view of the same threat. In addition, these frameworks oftentimes do not provide specific recommendations or initiatives and instead provide general ambiguous metrics from which specific initiatives have to be inferred. For example, ambiguous metrics include generic measures of ``high,'' ``medium,'' or ``low'' representing the probability of being compromised and the severity of impact for a broad organizational domain. While addressing these shortcomings, the subsequent framework synthesizes and adapts the strengths of previous frameworks while centering itself around an IoT-specific attack taxonomy that can be leveraged for automated analysis.

\subsection{IoT Threat Modeling}
\begin{figure}[h]
\centering
\includegraphics[width=\linewidth]{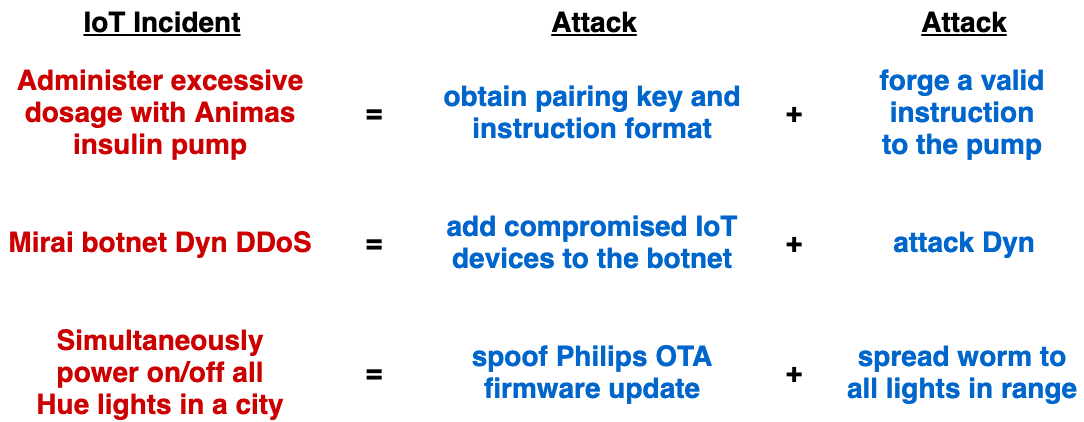}
\caption{Examples of Classifying IoT Incidents}
\label{fig:IoTIncidents}
\end{figure}

Due to the shortcomings that are present in these insufficient approaches, it is paramount that we establish new threat modeling and risk assessment frameworks that analyze security and privacy risks unique to IoT ecosystems. While previous work has assisted in modeling IoT threats \cite{IoTThreatTaxonomy,OWASPIoT}, we wish to provide a full framework of IoT threat modeling, risk assessment, and maturity assessment, where the risk and maturity assessments are informed by the IoT threat models. Our approach is to design and leverage an IoT attack taxonomy as the basis for threat modeling and risk assessment. The IoT attack taxonomy is comprised of a comprehensive list of individual attacks. Because we cannot suitably classify every IoT \emph{incident} into a taxonomy, we instead classify the individual \emph{attacks} that make up the incident. Each attack functions as another step towards effectively carrying out the IoT incident so that any IoT incident is composed of a series of individual attacks. Figure \ref{fig:IoTIncidents} provides a few examples of IoT incidents and the individual attacks that make up those incidents.
For example, the Mirai botnet that was used to carry out a distributed denial of service (DDoS) attack on Dyn can be decomposed into two parts: 1) adding compromised IoT devices to the botnet and 2) attacking Dyn. We classify each individual part of this attack in the IoT attack taxonomy since it is difficult to classify the entirety of the Mirai botnet incident into a taxonomy. Further examples for IoT incidents related to the Animas insulin pump and Philips Hue lights are also provided in Figure \ref{fig:IoTIncidents}.

\begin{figure}[h]
\centering
\includegraphics[width=\linewidth]{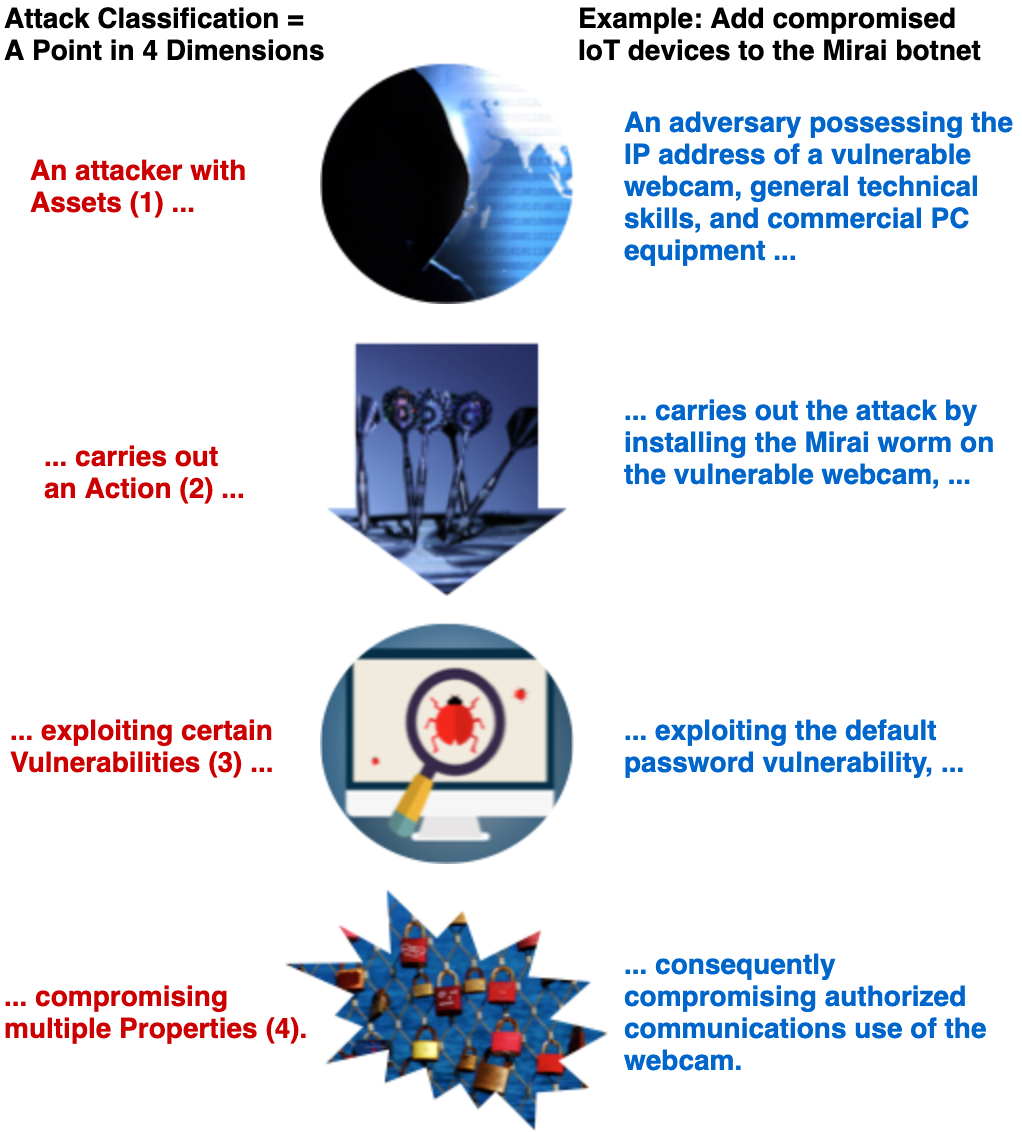}
\caption{Classifying IoT Attacks: Mirai Botnet}
\label{fig:AttackExample}
\end{figure}

As depicted in Figure \ref{fig:AttackExample}, any individual attack can be broken down into four parts, each of which is associated with a particular dimension of the IoT attack taxonomy: an attacker with a specific set of 1) \emph{assets} carries out a particular 2) \emph{action} to exploit one or more 3) \emph{vulnerabilities}, compromising a particular set of 4) \emph{properties}. For example, the individual attack of adding compromised IoT devices to the Mirai botnet can be broken down into 1) an adversary possessing the IP address of a vulnerable webcam, general technical skills, and commercial PC equipment 2) carries out the attack by installing the Mirai worm on the vulnerable webcam, 3) exploiting the default password vulnerability, 4) consequently compromising the authorized communications use of the webcam.
Breaking down the steps of an individual attack into these four dimensions allows us to chain together the individual attacks that make up an IoT incident through the asset dimension which includes prior information and resources obtained from previous individual attacks. For example, in the second individual attack of the Mirai botnet IoT incident, the attacker leverages the resources obtained in the previous individual attack, namely the authorized communications use of the webcam, to attack Dyn.

By using an IoT attack taxonomy as the foundation for threat modeling and risk assessment, we can comprehensively characterize and break down the complex interactions between IoT devices and cover the entire known IoT attack space. The IoT attack taxonomy also accounts for the unfixable flaws and diversity prevalent in the IoT in addition to the IoT's direct relationship to physical processes. Consequently, the taxonomy explicitly accounts for the unique factors that distinguish the IoT in its susceptibility to various types of attacks.

\begin{figure}[h]
\centering
\includegraphics[width=\linewidth]{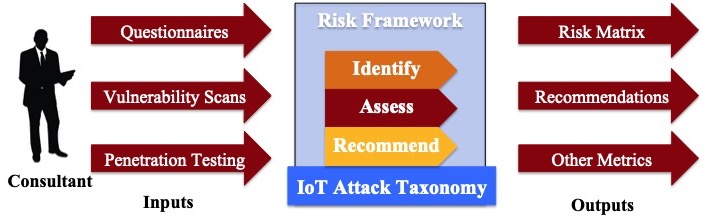}
\caption{IoT Attack Taxonomy Automated Assessment Methodology}
\label{fig:Consultant2}
\end{figure}

In addition, using an IoT attack taxonomy as the foundation for threat modeling and risk assessment enables the framework to be quantitative, reducing the variability that comes as a result of heavily depending on the consultant. As seen in Figure \ref{fig:Consultant2}, the IoT attack taxonomy enables the consultant to simply provide inputs and analyze outputs of the automated framework instead of being heavily involved in the framework itself as depicted in Figure \ref{fig:Consultant1}. This foundation also allows the framework to easily adapt to the discovery of new attacks by simply adding or removing elements and relationships between elements in the taxonomy.

Furthermore, the IoT attack taxonomy meets the criteria necessary for developing effective taxonomies \cite{Taxonomy}. This taxonomy is 1) \emph{mutually exclusive} in the sense that each attack can only be classified into one category since the categories do not overlap. The IoT attack taxonomy is also 2) \emph{complete} and \emph{exhaustive} in that the provided categories account for all possible known attacks. In addition, this taxonomy is 3) \emph{comprehensible} in that it contains precise, clear, and concise information so that any terminology and classification are not uncertain or confusing. Lastly, the IoT attack taxonomy is 4) \emph{useful} by providing logical and intuitive categories that contribute insight to the IoT attack space, supporting easy classification of new attacks and allowing the addition of newly developed attack categories. A detailed view of the IoT attack taxonomy is available at \url{http://iotrisk.andrew.cmu.edu}.\vspace{0.1cm}

\noindent\textit{1) Attacker Assets}\\\indent
The first dimension of the IoT attack taxonomy enumerates the assets, capabilities, and capacity needed by an attacker to carry out an IoT attack. This dimension is composed of six elements, each of which describes a different aspect of the attacker's capabilities. Many IoT-specific attack capabilities are included in this dimension such as possessing an IoT botnet and being able to manipulate particular sensors. The six elements of this dimension are modeled after the standards presented in \cite{NISTRiskAssessment,ISO27005,NIST800-37,NIST800-39,ISO31000} and inform how likely it is for an IoT attack to be attempted. These six elements are described as follows.

The 1) \emph{prior information} gathered by an attacker refers to the prerequisite knowledge needed to carry out an IoT attack. This information can be publicly available, gathered through surveillance, supplied by an insider, or stolen information. The 2) \emph{location} or \emph{access} of an attacker refers to where the adversary is when carrying out an IoT attack. This location can be remote, within wireless range, on the same network, or have physical access. The 3) \emph{equipment} used by an attacker may refer to commercial hardware equipment, a distributed system, or specialized equipment or facilities that are necessary in order to carry out an attack. The 4) \emph{technical skills} required by an adversary to carry out an attack can include basic skills, general programming skills, specific niche skills, or multiple advanced specific niche skills. The 5) \emph{time requirement} refers to the duration of time needed to successfully carry out an attack, whether that be a short period, long continuous period, or a specific time slot. Lastly, the 6) \emph{persistence requirements} refer to how much an adversary needs to maintain a presence in order for an attack to be successful, and this can range from nothing to being able to adapt the attack to avoid detection.\vspace{0.1cm}

\noindent\textit{2) Attacker Actions}\\\indent
The second dimension of the IoT attack taxonomy describes the actions and mechanisms an attacker would use to exploit vulnerabilities, and it is modeled after the standards in \cite{NISTRiskAssessment,ISO27005,NIST800-37,NIST800-39,ISO31000}. Each action is largely independent from both the locus of the vulnerability and the consequence of the exploit. The highest taxonomic level in this dimension refers to the attack mechanism while the middle and lower levels of classification are given by the attack pattern category and the attack pattern itself, respectively. This dimension is composed of seven different categories of attack mechanisms, which include 1) \emph{collecting and analyzing information}, 2) \emph{employing probabilistic techniques}, 3) \emph{engaging in deceptive interactions}, 4) \emph{manipulating data structures}, 5) \emph{abusing existing functionality}, 6) \emph{subverting access control}, and 7) \emph{manipulating system resources}. Each of these categories includes many IoT-specific actions such as tag tracking; node replication, spoofing, and injection; replay attacks; bypassing physical security; contaminating the physical environment; and hardware tampering.\vspace{0.1cm}

\noindent\textit{3) Exploitable Vulnerabilities}\\\indent
The third dimension of the IoT attack taxonomy is the dimension that the threat modeling and risk assessment frameworks are centered around. This dimension enumerates all the possible vulnerabilities that can be exploited by an attacker, classified according to vulnerabilities in 1) \emph{communications}, 2) \emph{software}, and 3) \emph{hardware} as suggested in \cite{NISTRiskAssessment,ISO27005,NIST800-37,NIST800-39,ISO31000}. Communications vulnerabilities are further divided into vulnerabilities relating to traffic encryption, traffic authenticity, privacy, protocol, and traffic obstruction, while software vulnerabilities are further divided into cloud control or storage interface vulnerabilities, embedded software vulnerabilities, and update mechanism vulnerabilities. Many of these software vulnerabilities are unique to the IoT such as unnecessary privileges and no integrity check during updates, and many of the hardware vulnerabilities are also IoT-specific including the lack of hardware tamper protection.\vspace{0.1cm}

\noindent\textit{4) Compromised Properties}\\\indent
The fourth dimension in the IoT attack taxonomy describes the properties that may be compromised and their associated consequences when any one of the vulnerabilities in the third dimension is exploited. This dimension is also derived from the standards presented in \cite{NISTRiskAssessment,ISO27005,NIST800-37,NIST800-39,ISO31000}. While the security properties of confidentiality, integrity, and availability are central to most frameworks, this dimension of the IoT attack taxonomy adds other relevant properties that are specific to the context of the IoT, such as physical safety. Furthermore, common properties such as integrity imply different meanings depending on the context under study. For example, integrity in the context of network communication may refer to a message not being spoofed, whereas integrity for a thermal sensor can refer to it reporting the correct temperature. The security properties identified in this dimension of the taxonomy are 1) \emph{confidentiality breaches}, 2) \emph{integrity breaches}, 3) \emph{losses of availability}, 4) \emph{authorization breaches}, and 5) \emph{safety breaches}.

\subsection{IoT Risk Domains}
In order to mitigate any threats or risks identified in the IoT attack taxonomy, appropriate security and privacy controls must be implemented. These controls can be classified according to the domain in which they mitigate risk. The risk domains used in this framework are designed and chosen such that all the vulnerabilities in the third dimension of the IoT attack taxonomy clearly fall within one of the risk domains. The risk domains, along with the specific controls that constitute those domains, are derived from standard frameworks including OWASP IoT Project \cite{OWASPIoT}, IoT Security and Privacy Trust Framework \cite{OTA}, ISO/IEC 27001 \cite{ISO27001}, NIST SP-800-53 \cite{NIST800-53}, ISO/IEC 27002 \cite{ISO27002}, and ISO/IEC 33004 \cite{ISO33004}. The nine risk domains identified are 1) \emph{governance and accountability}, 2) \emph{physical security}, 3) \emph{encryption}, 4) \emph{systems security}, 5) \emph{identity and access management}, 6) \emph{event logging and monitoring}, 7) \emph{supply chain security}, 8) \emph{threat and vulnerability management}, and 9) \emph{communications security}. Since the actions taken to mitigate risk differ significantly between producers and consumers, two sets of controls exist for each risk domain, one for producers and one for consumers. These controls include both technical controls and management controls. Unlike current frameworks, these controls include actions such as physical access control, maintenance, and disposal that address IoT-specific characteristics like physical security. A detailed view of these controls is available at \url{http://iotrisk.andrew.cmu.edu}.

\section{IoT Framework Overview}
\begin{figure}[h]
\centering
\includegraphics[width=\linewidth]{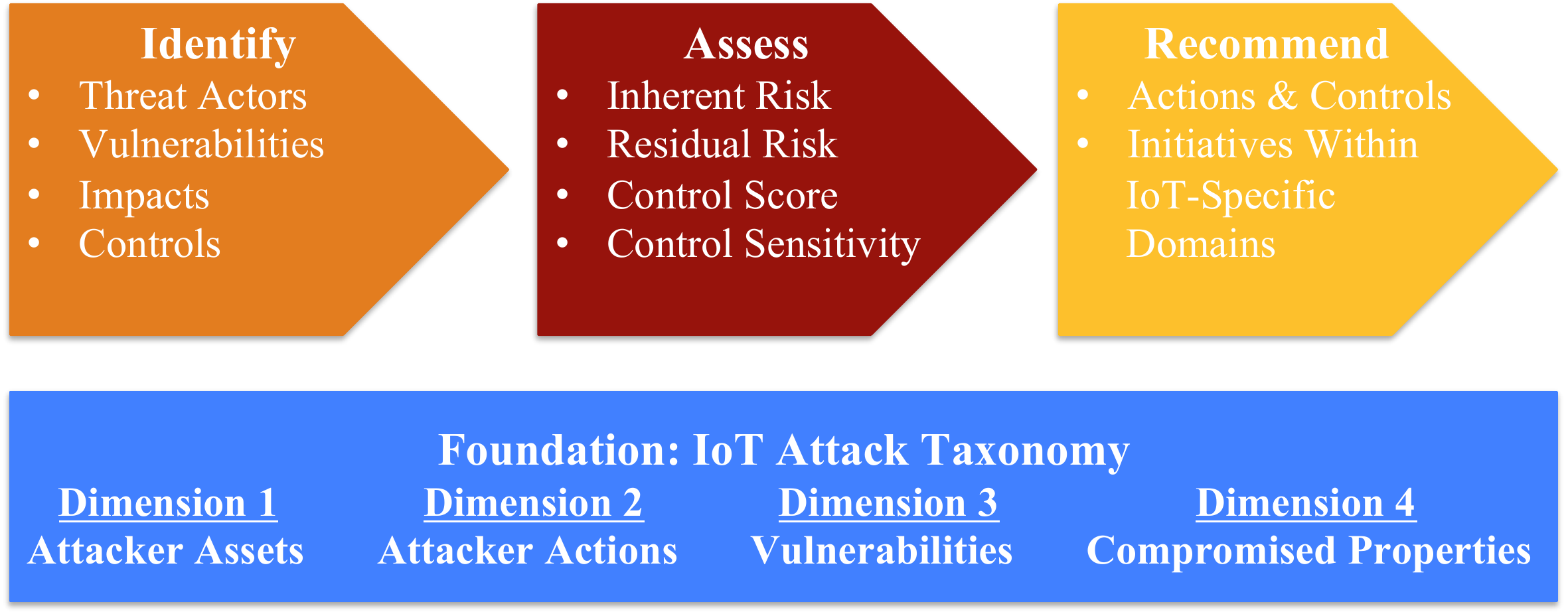}
\caption{IoT Framework Overview}
\label{fig:FrameworkOverview}
\end{figure}

Having presented the IoT attack taxonomy that serves as the basis for threat modeling and risk assessment, we now introduce the structure and overall approach of the interactive tool (\url{http://iotrisk.andrew.cmu.edu}) we developed to model threats and assess risks in the IoT. This overall approach can be broken down into three stages as summarized in Figure \ref{fig:FrameworkOverview}, all of which are founded on the IoT attack taxonomy.
The identify stage is comprised of a consultant gathering relevant information from the organization and using that information to inform which portions of the IoT attack taxonomy are relevant. Once this is completed, an automated and quantitative assessment stage is initiated which utilizes the information provided in the IoT taxonomy to compute a variety of metrics. These metrics are then used by the consultant in the recommend stage to provide specific suggestions and initiatives for the organization regarding its IoT ecosystems.

\subsection{Identify}
\begin{figure}[h]
\centering
\includegraphics[width=0.7\linewidth]{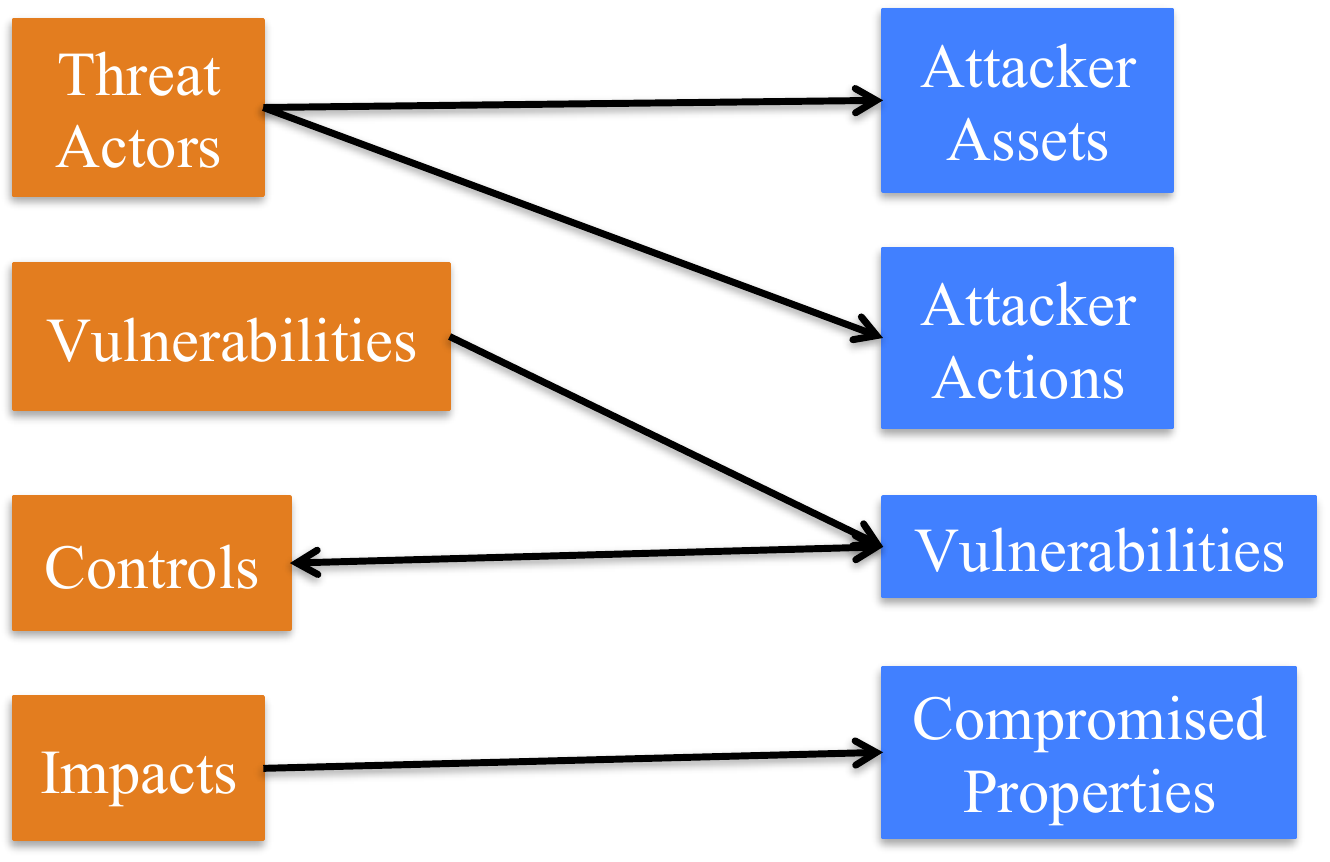}
\caption{Identifying Relevant Parts of the IoT Attack Taxonomy}
\label{fig:Identify}
\end{figure}

In the identify stage, a consultant compiles information about possible threat actors, potential vulnerabilities within the organization's IoT infrastructure, possible impacts suffered by the organization if it is attacked, and existing controls that the organization has in place to mitigate potential breaches. To ease the process of compiling this information, a few preset profiles such as threat actor profiles or IoT device profiles may be used. This stage is a central part of the threat modeling process. As depicted in Figure \ref{fig:Identify}, the potential threat actors inform the taxonomy regarding how likely it is for an adversary to have access to a particular asset or to carry out a particular action. The potential vulnerabilities and existing organizational controls both inform the taxonomy by indicating how prevalent specific vulnerabilities are within the organization. Lastly, the negative impacts that an organization might suffer help inform the extent to which the properties within the taxonomy could be compromised.

\subsection{Assess}
\begin{figure}[h]
\centering
\includegraphics[width=\linewidth]{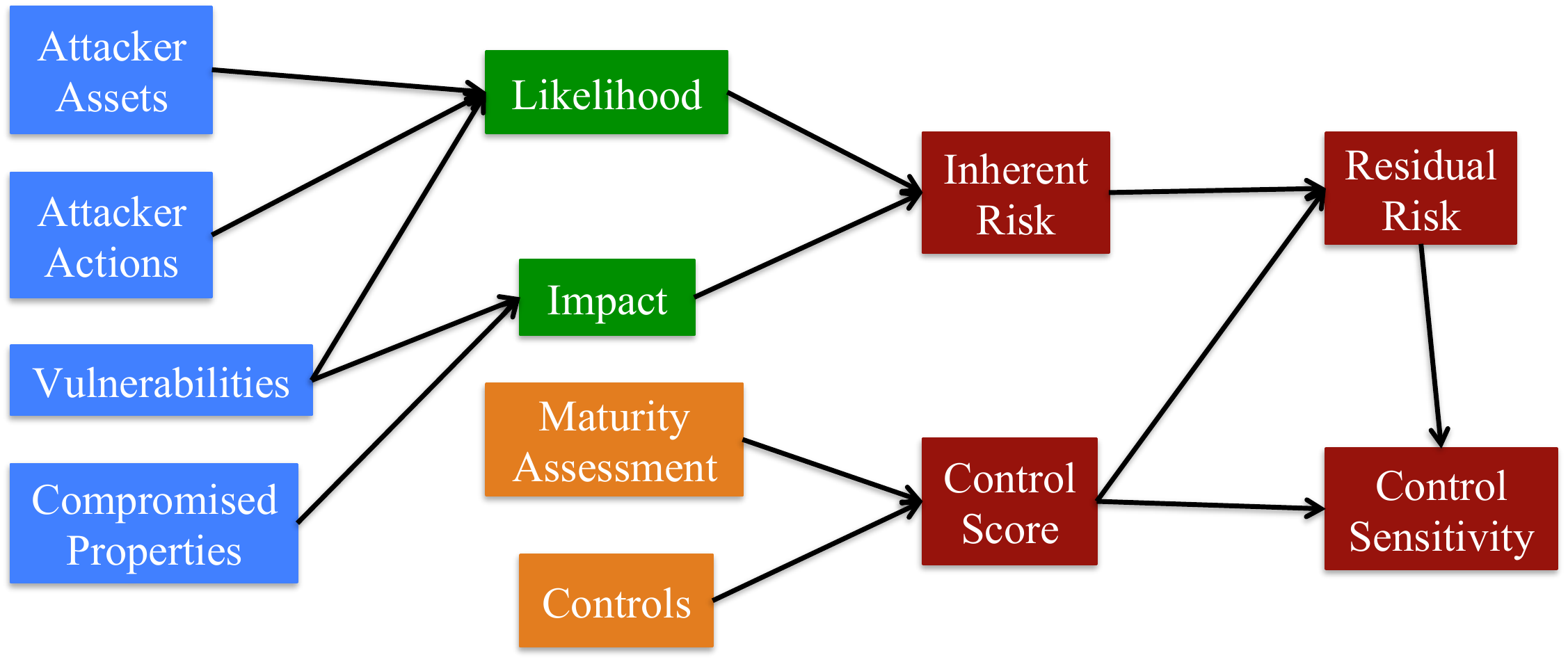}
\caption{Assessing Inherent Risk and Residual Risk}
\label{fig:Assess}
\end{figure}

The assessment stage, which is depicted in Figure \ref{fig:Assess}, can be divided into two substages: risk assessment and maturity assessment. During risk assessment, the four dimensions of the IoT attack taxonomy are leveraged to compute measures of likelihood and impact for each vulnerability, representing the likelihood that a vulnerability is exploited and the negative impact that would result if it were exploited. These measures of likelihood and impact are then used to compute measures of inherent risk within various domains of an organization. Inherent risk simply refers to the raw or untreated risk present in an organization before any actions are taken to reduce that risk. During maturity assessment, information that was gathered in the identification stage is used to assign measures of how effective and well-implemented each control is. These measures are then used to compute control scores that describe how well each control mitigates risk. The risk and maturity assessments are then used in tandem to compute measures of residual risk within various domains of an organization. Residual risk is simply the amount of risk remaining after an organization has implemented all of its risk-mitigating actions, procedures, and policies. This information is then used to conduct a sensitivity analysis that demonstrates which specific controls best mitigate risk.

\subsection{Recommend}
\begin{figure}[h]
\centering
\includegraphics[width=0.7\linewidth]{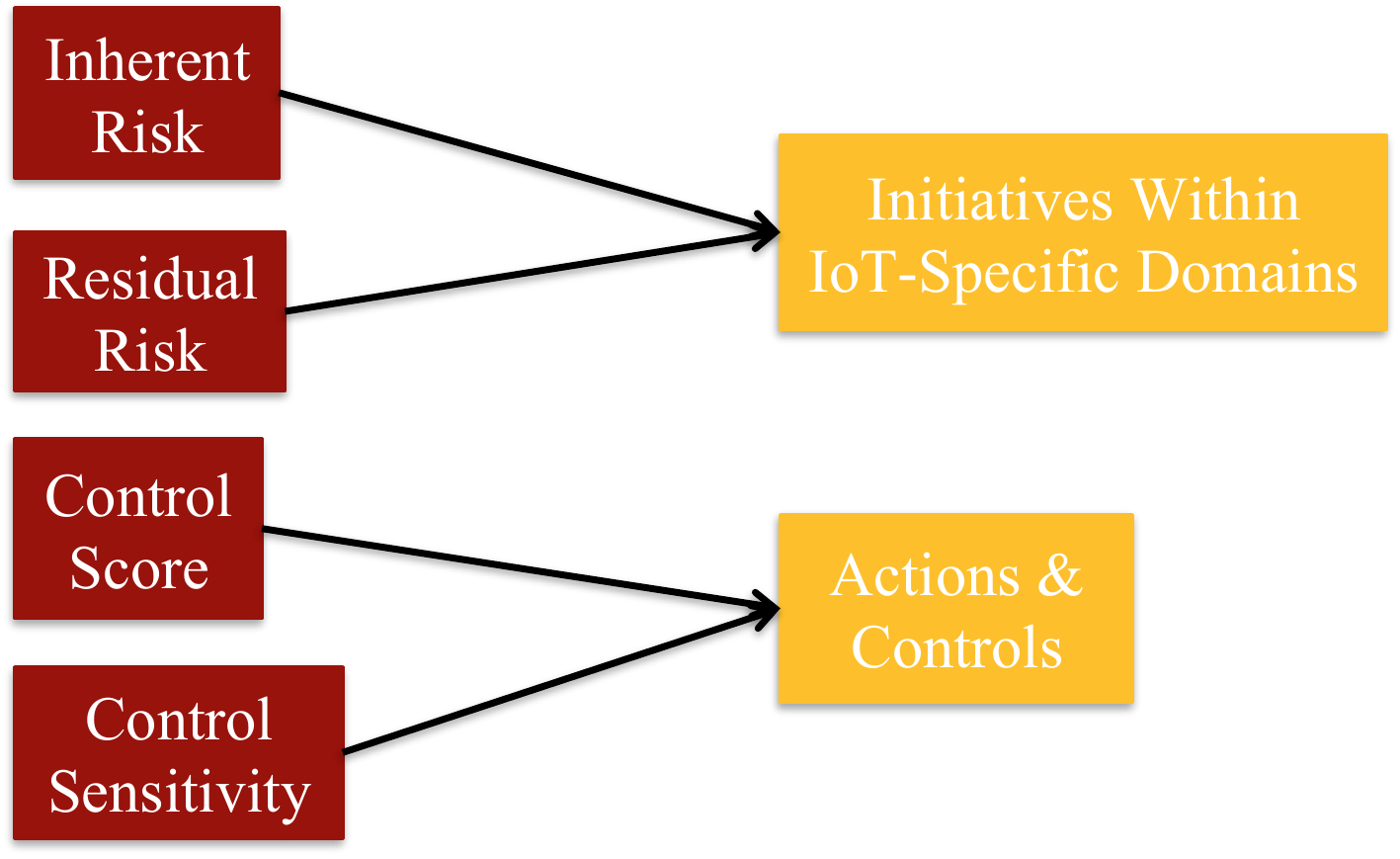}
\caption{Recommendations Based on Assessment}
\label{fig:Recommend}
\end{figure}

In the recommend stage, a consultant leverages the measures of inherent risk, residual risk, the control scores, and the sensitivity analysis to provide specific recommendations and initiatives to the organization about how to better mitigate risk within its various IoT environments. As seen in Figure \ref{fig:Recommend}, these recommendations include suggestions as to which actions and controls an organization can introduce or better implement to make its IoT environments more secure.

\section{Interactive Tool Implementation}
One of the main contributions of this paper lies in the development of an interactive tool (\url{http://iotrisk.andrew.cmu.edu}) that functions as the quantitative and automated assessment stage of the overall IoT framework. This section describes the contents of the interactive tool with respect to the implementation of the assessment stage and how this assessment leverages and grounds itself in the IoT attack taxonomy.

The assessment stage seeks to comprehensively characterize the threats an organization faces by leveraging the classifications given in the IoT attack taxonomy. Based on the information provided by CAPEC \cite{CAPEC} and OWASP \cite{OWASPIoT}, relationships or mappings between each dimension of the taxonomy are defined, providing a list of attacker assets that may be used to carry out specific attack actions, a list of attacker actions that may be used to exploit specific vulnerabilities, and a list of properties that are compromised when specific vulnerabilities are exploited. Using the IoT attack taxonomy and these relationships between each of its dimensions, measures of likelihood and impact can be calculated for each vulnerability, allowing a measure of inherent risk to be calculated for each IoT risk domain.

\subsection{IoT Risk Assessment}
To calculate measures of likelihood for each vulnerability, specific threat actor profiles are defined according to who might carry out an attack and their intention and motivation for carrying out the attack. Commonly chosen threat actor profiles include ideologically motivated threat actors such as a hacktivist, financially motivated threat actors, and state-sponsored threat actors such as a nation state.
\vspace{0.1cm}

\noindent\textit{1) Likelihood}\\\indent
To calculate the likelihood of a specific vulnerability being exploited, we consider the appropriate combinations of threat actors, attacker assets, and attacker actions that can be used to exploit the particular vulnerability. Using historical data and expert consensus, a likelihood score $p(s_i^j|t_k)$ is assigned to each asset $s_i^j$ given a specific threat actor $t_k$, representing the probability of threat actor $k$ possessing asset $i$ from sub-dimension $j$ of dimension 1. In addition, a likelihood score $p(a_i|t_j)$ is assigned to each action $a_i$ given a specific threat actor $t_j$, representing the probability of threat actor $j$ taking action $i$. The likelihood of the union of all possible combinations of threat actors, attacker assets, and attacker actions is used as the measure of likelihood for a particular vulnerability being exploited. By using this measure of likelihood, the IoT threat modeling framework comprehensively leverages all the information in the IoT attack taxonomy and all the relationships between its elements.

We first calculate the likelihood of a specific threat actor possessing at least one of the attacker assets in sub-dimension $j$ of dimension 1 that can be used to carry out action $k$. Using the inclusion-exclusion principle and assuming independence between attacker assets, this likelihood is given by
\begin{equation}
p\left(\bigcup_{s_i^j\in S_k^j} s_i^j\middle|t_n\right) = \sum_{\substack{i=1\\s_i^j\in S_k^j}}^{N_k^j}(-1)^{i-1}\sum_{\substack{I\subset\{1,\dots,N_k^j\}\\|I|=i}}\prod_{m\in I}p(s_m^j|t_n),
\end{equation}
where $S_k^j$ represents the set of assets from sub-dimension $j$ of dimension 1 that can be used to carry out action $k$, $N_k^j$ represents the number of elements in this set, and $|I|$ represents the cardinality of set $I$.

We then calculate the likelihood of a specific threat actor possessing at least one vector of assets $\vec{s}_q$ that can be used to carry out action $k$. Because dimension 1 is composed of six sub-dimensions describing an attacker's assets, asset vector $q$ refers to a set of six elements, one from each sub-dimension of dimension 1. Assuming independence between attacker assets, this likelihood is given by
\begin{equation}
p\left(\bigcup_{\vec{s}_q\in S_k} \vec{s}_q\middle|t_n\right) = \prod_{j=1}^6 p\left(\bigcup_{s_i^j\in S_k^j} s_i^j\middle|t_n\right),
\end{equation}
where $S_k$ represents the set of asset vectors that can be used to carry out action $k$.

Next, we calculate the likelihood of a specific vulnerability $v_r$ being exploited by a particular threat actor. Using the inclusion-exclusion principle and assuming independence between attacker assets and actions, we consider all possible combinations of asset vectors and actions that could be used to exploit the vulnerability. This likelihood is given by
\begin{equation}
\medmuskip=3.26mu
\thinmuskip=3.26mu
\thickmuskip=3.26mu
\begin{split}
& p(v_r|t_n) = p\left(\bigcup_{a_k\in A_r} \left(a_k,\bigcup_{\vec{s}_q\in S_k} \vec{s}_q\right)\middle|t_n\right) \\
& = \sum_{\substack{k=1\\a_k\in A_r}}^{N_r}(-1)^{k-1}\sum_{\substack{I\subset\{1,\dots,N_r\}\\|I|=k}}\prod_{m\in I}p(a_m|t_n)p\left(\bigcup_{\vec{s}_q\in S_m}\vec{s}_q\middle|t_n\right),
\end{split}
\end{equation}
where $A_r$ represents the set of actions that can be used to exploit vulnerability $r$ and $N_r$ represents the number of elements in this set.

Lastly, we calculate the likelihood $L_r$ of a specific vulnerability being exploited by at least one of the relevant threat actors. Using the inclusion-exclusion principle and assuming independence between threat actors, this likelihood is given by
\begin{equation}
\medmuskip=2.36mu
\thinmuskip=2.36mu
\thickmuskip=2.36mu
L_r = p\left(v_r\middle|\bigcup_{t_n\in T}t_n\right) = \sum_{\substack{n=1\\t_n\in T}}^N(-1)^{n-1}\sum_{\substack{I\subset\{1,\dots,N\}\\|I|=n}}\prod_{m\in I}p(v_r|t_m),
\end{equation}
where $T$ represents the relevant set of threat actors and $N$ represents the number of elements in this set.\vspace{0.1cm}

\noindent\textit{2) Impact}\\\indent
To calculate the impact associated with a particular vulnerability being exploited, we consider all the properties that are compromised when that vulnerability is exploited. Using historical data and expert consensus, an impact score $w_p^j$ is assigned to each property $p_j$, representing the monetary loss or damage incurred if property $j$ is compromised. The impact $I_i$ of exploiting vulnerability $i$ is simply calculated by adding together the associated impact scores so that
\begin{equation}
I_i = \sum_{p_j \in P_i} w_p^j,
\end{equation}
where $P_i$ represents the set of properties that are compromised when vulnerability $i$ is exploited.\vspace{0.1cm}

\noindent\textit{3) Inherent Risk}\\\indent
To calculate the inherent risk associated with a particular vulnerability, we use the standard equation for technology risk where the likelihood of exploiting the vulnerability is multiplied by the impact associated with exploiting the vulnerability \cite{OWASPRisk}. When calculating the inherent risk $R_i$ for risk domain $i$, we carry out a weighted sum of the inherent risks associated with each vulnerability in risk domain $i$ where the weights are the vulnerability prevalency scores. This computation is given by
\begin{equation}
R_i = \sum_{v_j \in V_i} p(v_j) L_j I_j,
\end{equation}
where $V_i$ represents the set of vulnerabilities associated with risk domain $i$.

To normalize the inherent risk so that all measures of inherent risk lie between a minimum value $r_{min}$ and a maximum value $r_{max}$, we scale the inherent risk by the maximum possible inherent risk. The maximum possible inherent risk occurs when all vulnerabilities are considered, the likelihood of exploiting any vulnerability is 100\%, and the impact score for every vulnerability is the maximum possible impact score. The normalized inherent risk $R_i^{norm}$ for risk domain $i$ is given by
\begin{equation}
R_i^{norm} = \frac{R_i(r_{max}-r_{min})}{\sum_{v_j \in V}p(v_j)I_j^{max}}+r_{min},
\end{equation}
where $V$ represents the set of all possible vulnerabilities and $I_j^{max}$ represents the maximum possible impact score for vulnerability $j$.

\subsection{IoT Maturity Assessment}
The IoT maturity assessment conveys an understanding of how well an organization has implemented risk-mitigating controls, allowing a measure of residual risk to be calculated for each IoT risk domain. Each control $c_i$ is assigned a control implementation score $p(c_i)$ and a control effectiveness score $p(e_i)$ that represent how well an organization has implemented that particular control and how effective that particular control is in mitigating risk, respectively. By interacting with the organization, these control scores can be assigned using answers to a maturity assessment questionnaire.\vspace{0.1cm}

\noindent\textit{1) Maturity Score}\\\indent
Using maturity scores, which represent the percentage reduction in inherent risk, a measure of residual risk can be calculated for each IoT risk domain. To calculate a maturity score for a specific vulnerability, we first calculate a mitigation percentage for each control where the mitigation percentage equals how well the control is implemented times how effective the control is. The mitigation percentage $p(m_i)$ for control $i$ is given by
\begin{equation}
\label{eq:mitigation}
p(m_i) = p(c_i)p(e_i)
\end{equation}
and represents how well control $i$ mitigates risk. Using the inclusion-exclusion principle and assuming independence between controls, the maturity score $M_j$ for vulnerability $j$ can be calculated as
\begin{equation}
\medmuskip=3.27mu
\thinmuskip=3.27mu
\thickmuskip=3.27mu
M_j = p\left(\bigcup_{m_i\in C_j} m_i\right) = \sum_{\substack{i=1\\m_i\in C_j}}^{N_j}(-1)^{i-1}\sum_{\substack{I\subset\{1,\dots,N_j\}\\|I|=i}}\prod_{k\in I}p(m_k),
\end{equation}
where $C_j$ represents the set of controls associated with vulnerability $j$ and $N_j$ represents the number of elements in this set.\vspace{0.1cm}

\noindent\textit{2) Residual Risk}\\\indent
To calculate the residual risk associated with vulnerability $j$, we multiply the inherent risk $L_jI_j$ associated with vulnerability $j$ times the unmitigated risk percentage $(1-M_j)$. When calculating the residual risk $Z_i$ for risk domain $i$, we carry out a weighted sum of the residual risk associated with each vulnerability in risk domain $i$ where the weights are the vulnerability prevalency scores. This computation is given by
\begin{equation}
Z_i = \sum_{v_j \in V_i}p(v_j)L_jI_j(1-M_j),
\end{equation}
where $V_i$ represents the set of vulnerabilities associated with risk domain $i$. The residual risk is normalized in the same way as the inherent risk so that
\begin{equation}
\label{eq:normRR}
Z_i^{norm} = \frac{Z_i(r_{max}-r_{min})}{\sum_{v_j \in V}p(v_j)I_j^{max}}+r_{min},
\end{equation}
where $Z_i^{norm}$ is the normalized residual risk for risk domain $i$.\vspace{0.1cm}

\noindent\textit{3) Sensitivity Analysis}\\\indent
After computing measures of inherent and residual risk, a sensitivity analysis is used to provide more detailed information about which specific controls should be better implemented to further reduce residual risk. In this framework, we conduct a one-at-a-time sensitivity analysis where a small value $\Delta p(c_j)$ is added to the control score $p(c_j)$ for control $j$, representing a small improvement in the implementation of control $j$. The normalized residual risk $Z_i^{norm}$ is recomputed as described in equations (\ref{eq:mitigation}) to (\ref{eq:normRR}), and a sensitivity score $\Delta Z_{ij}^{norm}$ for risk domain $i$ is assigned to control $j$. The sensitivity score $\Delta Z_{ij}^{norm}$ is simply the difference between the original residual risk and the recomputed residual risk. This process is repeated for each control $j$ in each risk domain $i$ until all sensitivity scores $\Delta Z_{ij}^{norm}$ have been assigned. The sensitivity scores in risk domain $i$ are then ordered by decreasing magnitude to show which controls have the largest effect on further reducing residual risk in that domain.

\section{Industrial Manufacturing Case Studies}

\subsection{Company X}
To test the accuracy and effectiveness of the proposed threat modeling, risk assessment, and maturity assessment framework, we applied the framework to analyze an anonymous Company X. Company X is a small industrial manufacturing company located in the midwestern United States with 5,000-10,000 employees and approximately \$2 billion in turnover. The company engages in the design, manufacture, and distribution of small gasoline engines and outdoor powered equipment including residential and commercial products. This company has invested in IoT technology for many of its products, including lawn mowers, irrigation controllers, door locks, robot vacuums, turf mowers, concrete mixers, drones, and tractors.

Company X is focusing on integrating connectivity amongst its devices to create an ecosystem enabling seamless communication. Existing products have been improved such as lawn mowers which trim grass by sensing the soil and the grass type and irrigation controllers which are controlled remotely or are scheduled to go off at certain times of the day. Mechanical door locks are equipped with sensors to open and lock doors based on the user's remote request via a smartphone. A cloud-based database hosts and manages all the data generated from these devices and is available on the go for consumers with real time reports indicating usage and ways to maximize efficiency.

The company is also striving to grow its connected industrial devices, including turf mowers, tractors, concrete mixers, and drones. These devices are equipped with sensors which assist in sending maintenance reminders directly to customers via SMS or email based on usage of the equipment. Real-time GPS tracking is employed to provide the exact location and status of the equipment. Geofences are deployed to restrict the equipment to operate in predefined areas. Clients can access detailed information on the engine to troubleshoot any mechanical or software issues. Data generated from the equipment is stored in a centralized database which is extracted for reports including total travel time, duration of each stop, and routes taken by the crew.

Company X is concerned with the availability and security of its key manufacturing systems, including any cyber event that would disrupt daily operations, cause loss of intellectual property, or cause an event affecting human safety. The company is also concerned about the reliability of its systems and products, including any failure of systems that would hinder internal daily operations or failure of products that would cause consumer frustration. In addition, the company is concerned with any financial loss it might potentially suffer, including anything that would hinder product development, sales, or tarnish its reputation.

Consequently, our framework was used to provide an appropriate analysis of the company's risks and controls with respect to its IoT technology. In this analysis, the scores for each element of the IoT attack taxonomy and the scores for each control were assigned in conjunction with expert consultants. They leveraged their experience and expertise, inventory of threat actor profiles, and previous IoT incidents to assign scores that captured the relational patterns between each dimension of the taxonomy in the context of industrial manufacturing and the IoT.

\begin{figure}[h]
\centering
\includegraphics[width=0.989\linewidth]{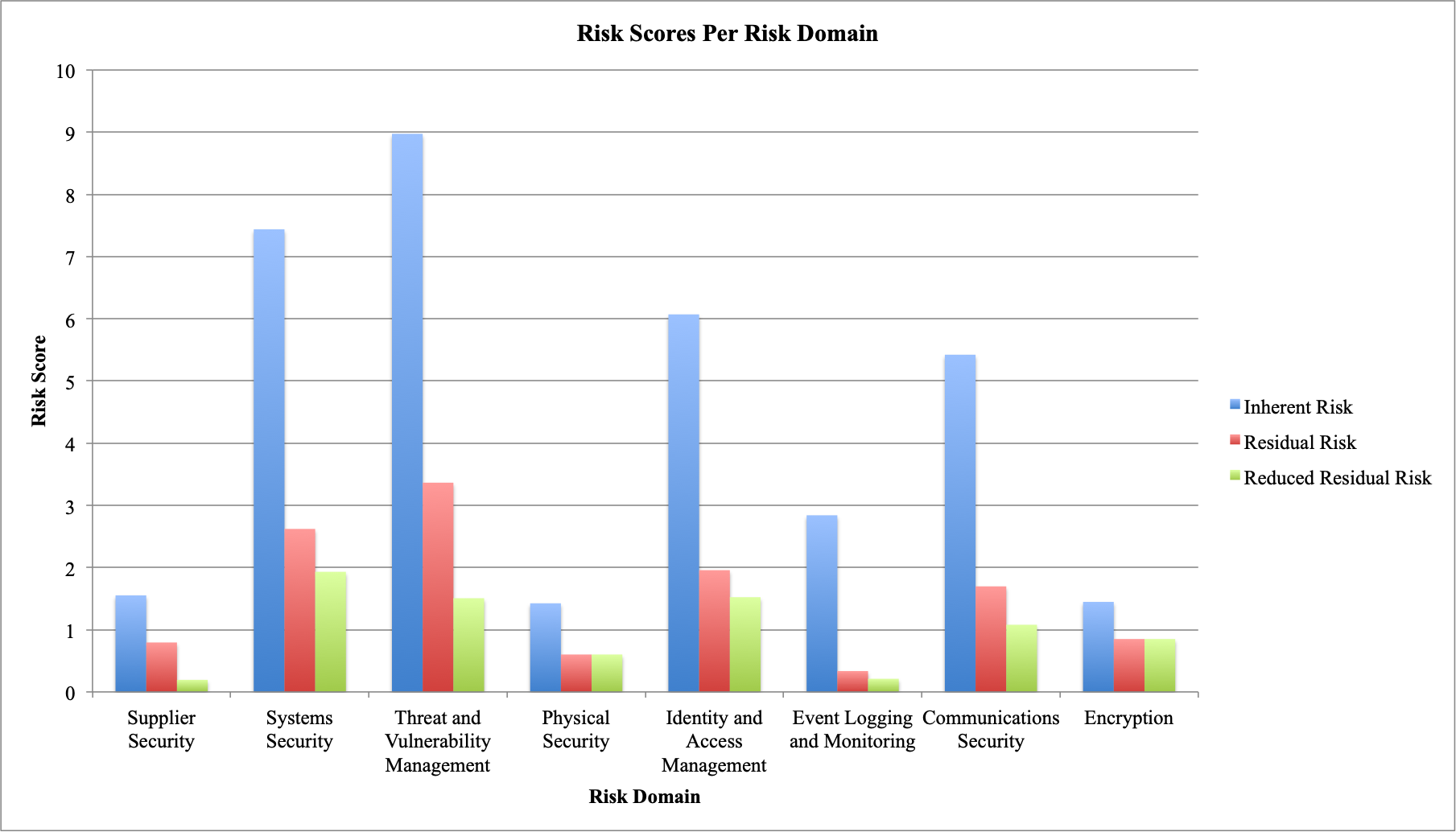}
\caption{Risk Scores Per Risk Domain for Company X}
\label{fig:Risks}
\end{figure}

Once these scores were assigned, we applied our framework to compute measures of inherent and residual risk in each risk domain of Company X as shown in Figure \ref{fig:Risks}. As can be seen, systems security and threat and vulnerability management are both relatively high areas of inherent risk for Company X. This is expected since Company X is concerned about human health and safety in addition to protecting the intellectual property in their devices. Figure \ref{fig:Risks} also shows that inherent risk has been reduced the most in event logging and monitoring and communications security, which is expected since Company X has implemented relatively mature controls in these domains. In general, the results presented in Figure \ref{fig:Risks} match the levels of risk the expert consultants expected to see.

\begin{figure}
\centering
\includegraphics[width=\linewidth]{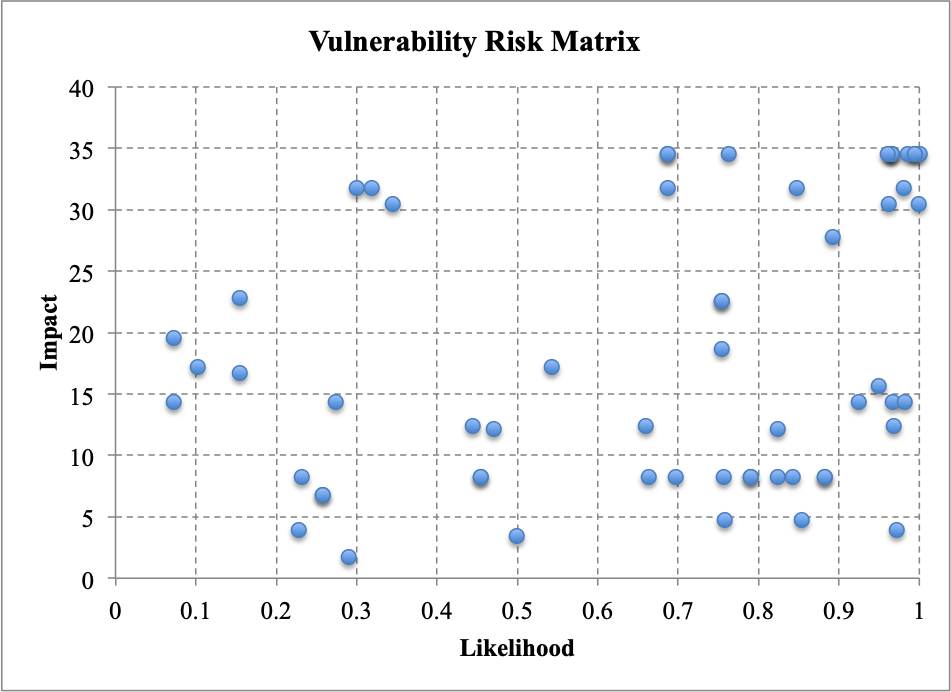}
\caption{Vulnerability Risk Matrix for Company X}
\label{fig:VRMatrix}
\end{figure}

\begin{figure}[h]
\centering
\includegraphics[width=0.989\linewidth]{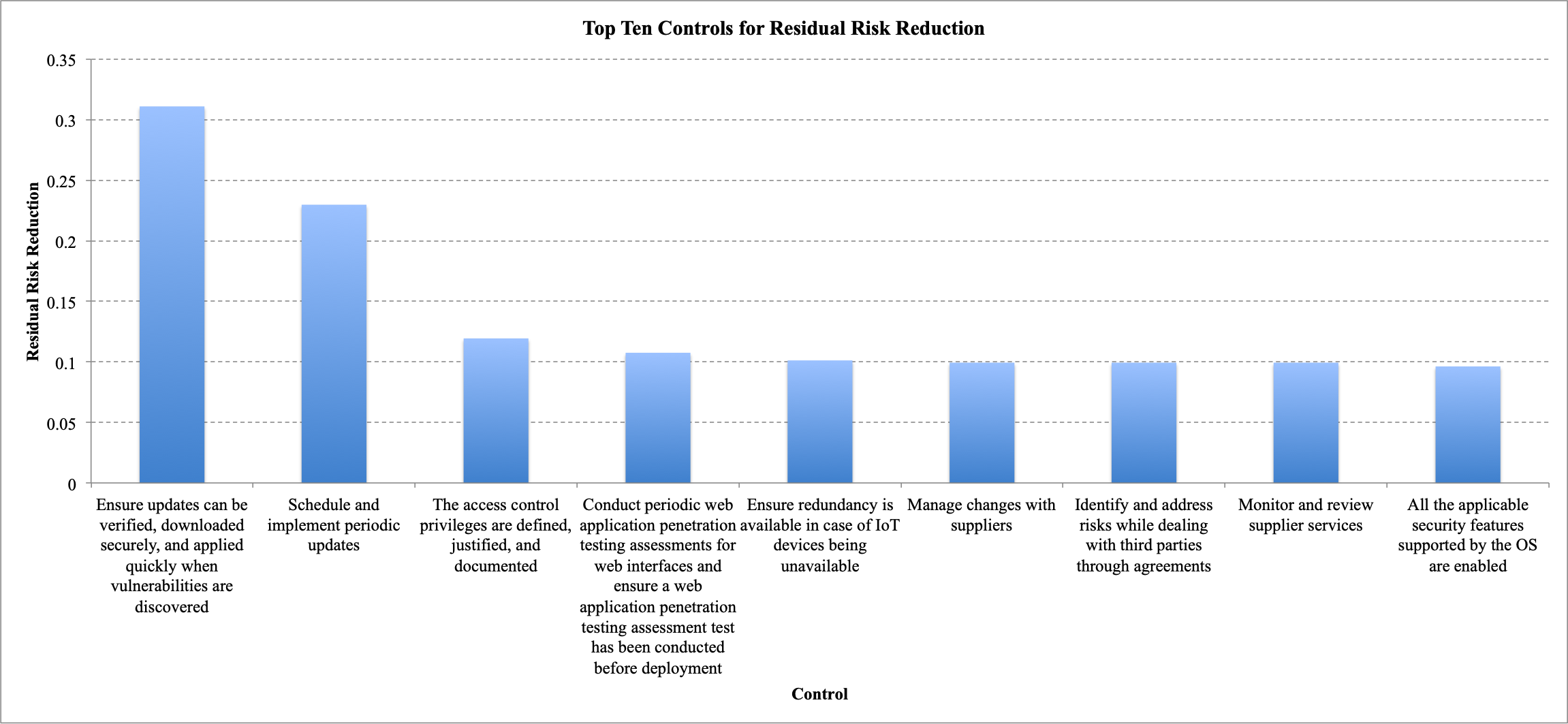}
\caption{Top Ten Controls for Residual Risk Reduction for Company X}
\label{fig:Sensitivities}
\end{figure}

Figure \ref{fig:VRMatrix} presents a scatter plot depicting the likelihood and impact scores associated with each of Company X's vulnerabilities, helping Company X decide which vulnerabilities to target when mitigating risk, namely those with high likelihood and impact. Figure \ref{fig:Sensitivities} shows the reduction in residual risk associated with further implementing each control by 10\%, assisting Company X in deciding which controls to implement and what actions to take to further mitigate risk. Figure \ref{fig:Risks} depicts the reduced residual risk resulting from increasing the implementation of these ten controls suggested in Figure \ref{fig:Sensitivities} by 30\%. As can be seen, the suggestions given in Figure \ref{fig:Sensitivities} are effective in reducing Company X's residual risk in all risk domains, including a significant reduction in threat and vulnerability management. The results displayed in these plots generally match the recommendations of the expert consultants for reducing Company X's risk, which include technical controls for device hardening within the IoT ecosystem in addition to implementing more robust measures for updating and patching. Our framework also provides the suggestions given in Figure \ref{fig:Sensitivities} for each risk domain, providing Company X with recommendations about further mitigating risk for specific domains.

\subsection{Company Y}
\begin{figure}[h]
\centering
\includegraphics[width=\linewidth]{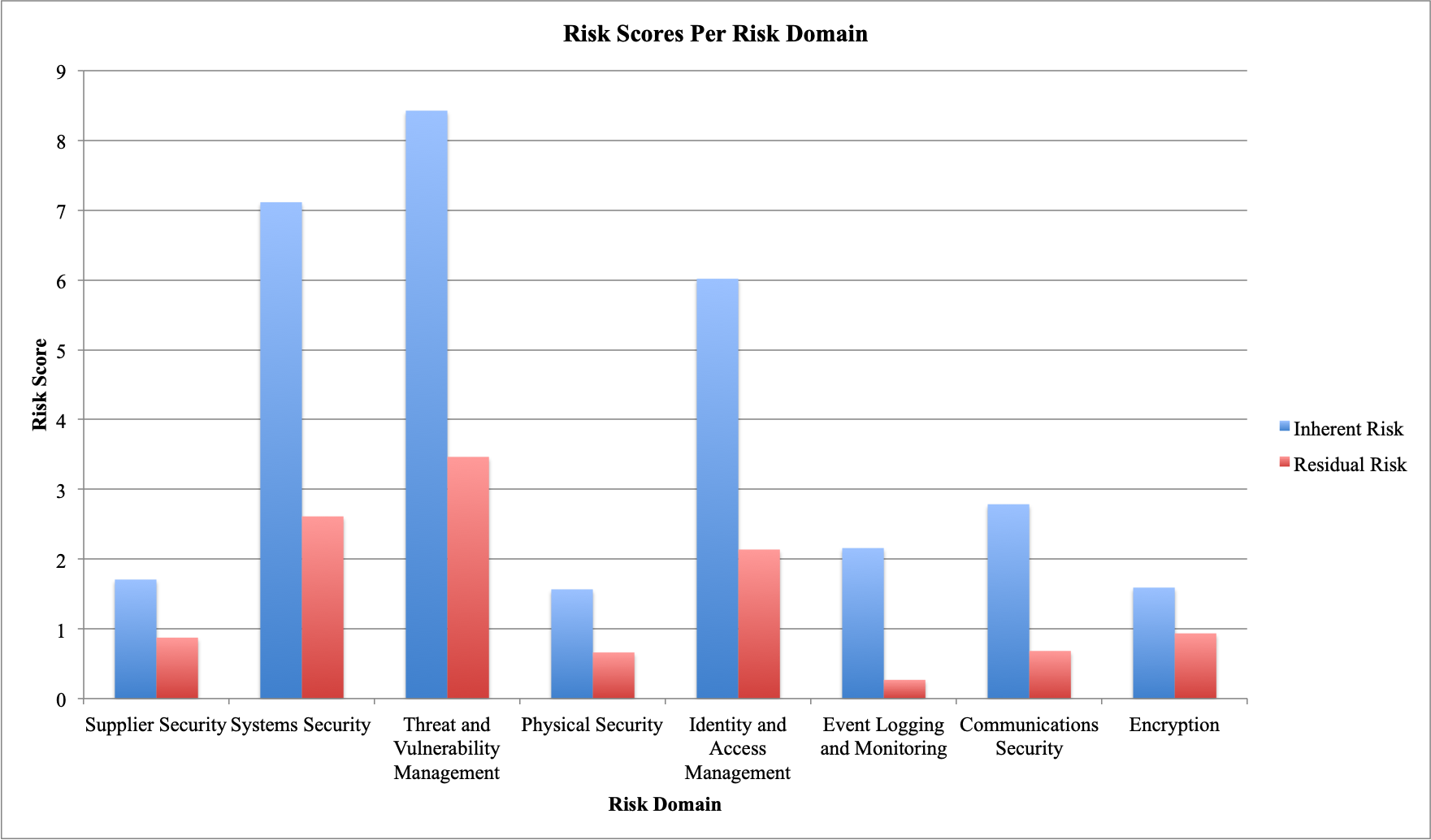}
\caption{Risk Scores Per Risk Domain for Company Y}
\label{fig:Risks2}
\end{figure}

To demonstrate the accuracy of our framework in computing measures of risk, we applied our framework to Company Y and compared the results to those obtained from Company X. Company Y is identical to Company X except that Company Y uses devices that exhibit many fewer communications vulnerabilities such as cleartext authentication and vulnerable SSL \cite{SoKIoT,SmartIoT}. Consequently, our framework should supply a much lower inherent risk score for the communications security risk domain for Company Y than it does for Company X. Figure \ref{fig:Risks2} displays the inherent and residual risk for each risk domain of Company Y. As expected, the inherent risk for communications security is significantly lower than that shown for Company X in Figure \ref{fig:Risks}, confirming the accuracy of the framework. An interactive implementation of this framework is available at \url{http://iotrisk.andrew.cmu.edu}.

\section{Conclusion}
With the continued growth of the IoT, organizations must begin assessing threats and risks in IoT environments. To do so, we have synthesized previous approaches to develop an IoT threat modeling framework. This framework utilizes an IoT attack taxonomy that describes the adversarial assets, adversarial actions, exploitable vulnerabilities, and compromised properties that are components of any IoT attack. We then leverage this taxonomy as the foundation for a risk assessment framework that provides a quantifiable measure of inherent risk in specific risk domains of an organization. We have also developed a maturity assessment framework that leverages information from the threat modeling framework to provide a measure of residual risk in the same risk domains. This framework provides organizations with specific recommendations as to where resources should be devoted to further mitigate risk. Rather than being a framework that relies heavily upon the subjective analysis of the consultant, this framework is more structured, quantitative, and consistent. It allows the consultant to focus on obtaining relevant information necessary for

\onecolumn

\newpage

\begin{multicols}{2}
\noindent assessment, ensuring that information is as accurate as possible, and analyzing the outputs provided by the framework. This framework still allows for flexibility through modifications of the taxonomy but is no longer directly dependent upon the variability between consultants. We have demonstrated the effectiveness of this IoT framework by implementing it in the context of multiple industrial manufacturing companies to provide recommendations for mitigating risk.

\section*{Acknowledgments}
The researchers gratefully acknowledge the support of the Risk and Regulatory Services Innovation Center at Carnegie Mellon University sponsored by PwC. We also acknowledge Alexandra Snoy and Gireesh Devadas for helping develop the IoT framework in addition to Siddhant Jain, Vishwas Singh, and Kshitij Gurjar for putting together the online interactive tool. Lastly, we thank Vyas Sekar for his valuable comments on early versions of this paper.

\bibliographystyle{IEEEtran}
\bibliography{root}
\end{multicols}

\end{document}